\documentclass[%
 reprint,
 amsmath,amssymb,
 aps,
]{revtex4-2}

\usepackage{graphicx}
\usepackage{dcolumn}
\usepackage{bm}
\usepackage{xcolor}
\usepackage[normalem]{ulem}

\begin{document}


\title{Isochronous island bifurcations driven by\\ resonant magnetic perturbations in Tokamaks}

\author{B. B. Leal}%
 \email{bruno.borges.leal@usp.br}

\author{I. L. Caldas}%
\affiliation{%
 Instituto de Física, Universidade de São Paulo, São Paulo, Brazil
}%

\author{M. C. de Sousa}
 
\affiliation{
 LSI, CEA/DRF/IRAMIS, CNRS, École Polytechnique, Institut Polytechnique de Paris, F-91128 Palaiseau, France
}%

\author{R. L. Viana}
\affiliation{%
 Departamento de Física, Universidade Federal do Paraná, Curitiba, Paraná, Brazil
}%

\author{A. M. Ozorio de Almeida}
\affiliation{Centro Brasileiro de Pesquisas Físicas, Rio de Janeiro, Brazil}

\date{\today}

\begin{abstract}
Recent evidences show that heteroclinic bifurcations in magnetic islands may be caused by the amplitude variation of resonant magnetic perturbations in tokamaks. To investigate the onset of these bifurcations, we consider a large aspect ratio tokamak with an ergodic limiter composed of two pairs of rings that create external primary perturbations with two sets of wave numbers. An individual pair produces hyperbolic and elliptic periodic points, and its associated islands, that are consistent with the Poincaré-Birkhoff fixed point theorem. However, for two pairs producing external perturbations resonant on the same rational surface, we show that different configurations of isochronous island chains may appear on phase space according to the amplitude of the electric currents in each pair of the ergodic limiter. When one of the electric currents increases, isochronous bifurcations take place and new islands are created with the same winding number as the preceding islands. We present examples of bifurcation sequences displaying (a) direct transitions from the island chain configuration generated by one of the pairs to the configuration produced by the other pair, and (b) transitions with intermediate configurations produced by the limiter pairs coupling. Furthermore, we identify shearless bifurcations inside some isochronous islands, originating nonmonotonic local winding number profiles with associated shearless invariant curves.

\end{abstract}

\maketitle


\section{\label{sec:level1} Introduction}

Adjustment of the magnetic field at the plasma edge is essential to improve the plasma confinement in tokamaks \cite{horton2015, horton2017}. To this purpose, the equilibrium field in this region is usually modified by resonant perturbations produced by electric currents in external coils. These external perturbations induce magnetic islands on rational magnetic surfaces where the safety factor assumes the value $q =m/n$, with the integers $m$ and $n$ the poloidal and toroidal wave numbers respectively \cite{hazeltine2003}. In tokamaks, the presence of these islands plays an important role in the plasma confinement, stability, and particle transport \cite{evans2013}. Thus, magnetic islands can improve or degrade the plasma performance.

Much work has been dedicated to controlling or modifying the magnetic islands. To minimize the damage caused by natural islands, several restrictions to the main plasma parameters have been proposed \cite{horton2017}. On the other hand, several devices have been used in tokamaks to modify or create magnetic islands at the plasma edge, altering the magnetic configuration and improving the plasma confinement in different ways \cite{schaffer2008, evans2008}. For example, in Pulsator and TBR, resonant helical windings were applied to avoid plasma disruptions \cite{karger1977, vannucci1988}; in Textor \cite{finken1997}, TextUp \cite{mccool1989} and Tore Supra \cite{ghendrih2002}, ergodic magnetic limiters were used to control plasma turbulence; in TCABR an ergodic limiter was used for Magnetohydrodynamics (MHD) mode control \cite{pires2005}. Later, in several tokamaks as in DIII-D \cite{boedo2001}, JET \cite{nardon2009}, ADEXP \cite{faitsch2018} and NSTX-U \cite{wu2019}, external coils have been used to control Edge Localized Modes (ELMs).

Recently, attention has been given to experiments with natural and external modes with different wave numbers and the same winding number, thus resonant at the same rational magnetic surface. For the DIII-D \cite{evans2021} and NSTX-U \cite{wu2019} tokamaks, resistive MHD simulations with external resonant magnetic perturbations predict heteroclinic bifurcations among isochronous islands (i.e. islands with the same winding number \cite{egydio1992,voyatzis2002,Sousa_2013,sousa2014,sousa2015}). More precisely, as the amplitude of an externally applied magnetic field perturbation is increased, the resonant helical magnetic islands are altered and undergo heteroclinic bifurcations. This kind of bifurcation has been observed in an experiment performed at the DIII-D tokamak \cite{bardoczi2021}. However, in the cited articles, the dynamics of the reported heteroclinic bifurcations were not described, motivating the investigation presented in this article.

Isochronous island chains have also been recognized in other areas of physics, such as molecular physics \cite{egydio1992}, periodic lattices \cite{lazarotto2022}, nonlinear oscillators \cite{sousa2015}, and electron beam interactions with electrostatic waves \cite{Sousa_2013,sousa2015}. For the case of wave-particle interactions, isochronous island chains were reported in a nearly integrable model \cite{Sousa_2013,sousa2015} describing the dynamics of relativistic charged particles moving in a uniform magnetic field and kicked by standing electrostatic pulses. Different island chains were identified for perturbative terms with the same winding number that generate isochronous islands in the same region of phase space. In the present article, our bifurcation analysis is similar to the one presented in \cite{Sousa_2013,sousa2014,sousa2015}.

To investigate the onset of the reported heteroclinic bifurcations of isochronous islands in tokamaks, one can consider perturbing magnetic fields generated by different types of external coils and/or determined by different models that are used to integrate the differential field line equation describing the perturbed field lines. One example is the resonant magnetic perturbation due to coils as those applied in TextUp \cite{mccool1989}, Textor \cite{finken1997} or DIII-D \cite{boedo2001}.

In this article, we consider the perturbing magnetic field due to an ergodic limiter, which consists of grid-shaped coils of finite width wound around the torus in the poloidal direction \cite{karger1977, pires2005}. The resonant magnetic field generated by this limiter creates magnetic island chains on rational magnetic surfaces at the plasma edge. Due to the fast radial decrease of the perturbation field, only the peripheral islands have a significant width. Therefore, the inner region of the plasma column is not much affected by an ergodic magnetic limiter.

We also consider a large aspect ratio tokamak with an ergodic limiter consisting of two pairs of rings, one pair exciting a mode with wave numbers $m_1$ and $n_1$, and the other one with wave numbers $m_2$ and $n_2$. Both modes present the same winding number $\iota = n_1 / m_1 = n_2 / m_2$, resonant at the same rational surface with safety factor $q = 1 / \iota$. 
 
An important technique that allows the integration of field lines for long times is to replace the field line trajectories with those of a symplectic map, i.e. a map that has the same Hamiltonian structure as the field lines \cite{portela2008, abdullaev_2014_book}. Individual trajectories obtained from the map can differ quantitatively from those obtained from the differential equations, but maps tend to depict correctly the qualitative dynamics. 

Here, the numerical results are obtained by using a two-dimensional symplectic map to describe the ergodic limiter action: a modified version of the Ullmann map \cite{ullmann2000}. Other symplectic maps have also been used in the tokamak literature. See, for example, the reviews in Refs. \cite{portela2008, abdullaev_2014_book}. Furthermore, in Ref. \cite{abdullaev_2014_book} procedures to derive these maps are presented. Some symplectic maps are good for exploring divertor dynamic properties \cite{punjabi1992, punjabi2003}, others to explore bifurcations or destruction of magnetic surfaces at the plasma edge \cite{martin1984, balescu1998, abdullaev2011, mugnaine2023}. Any of these maps could be used in our investigations. However, the Ullmann map has the advantage of allowing the choice of equilibrium profiles and perturbations with control parameters related to real values usually observed in experiments. The plasma screening and the plasma rotation are not included though, as in other maps.

The Ullmann map determines the crossings of the magnetic field lines with a surface defined by a transverse cross-section of the tokamak toroidal chamber. The two perturbative modes $(m_1, n_1)$ and $(m_2, n_2)$ are resonant on the same magnetic surface. One of the limiter pairs is held at a constant electric current, while the electric current in the other pair is varied. The two induced modes compete with each other so that there are only islands of one of these modes or of intermediate modes. As the electric current of the second pair varies, we observe saddle-node isochronous bifurcations with the creation of sets of hyperbolic-elliptic, i.e. unstable-stable, periodic points consistent with the Poincaré-Birkhoff fixed point theorem \cite{birkhoff1927,berry1978,lichtenberg1992,voyatzis2002}.

Furthermore, peculiar modifications in the magnetic configuration may appear. In particular, we find a shearless bifurcation inside the islands for which the internal winding number profile changes from monotonic to nonmonotonic. In this case, we also detect the formation of secondary isochronous twin island chains around a shearless curve inside the main primary islands. 

The paper is organized as follows. In Section \ref{Magnetic_Field}, we present the equilibrium magnetic configuration needed to confine the plasma, and how it is affected by the external perturbation produced by an ergodic magnetic limiter. We also discuss the winding number that determines the rational or irrational nature of each magnetic surface. In Section \ref{Bmapping}, we derive the modified Ullmann map version for a limiter with two pairs of rings. Using this map, it is possible to simultaneously obtain two different perturbations at the same magnetic surface.
In Sections \ref{Heteroclinic_Bifurcations} and \ref{Shearless_Bifurcations}, we present the results on isochronous heteroclinic bifurcations and shearless bifurcations obtained with the modified Ullmann map derived in Section \ref{Bmapping}. Finally, we draw our conclusions in Section \ref{Conclusions}.

\section{Magnetic Field}
\label{Magnetic_Field}

A plasma in a tokamak is confined by a magnetic configuration mainly composed of the superposition of two magnetic fields. One of them is the poloidal magnetic field created by the plasma current, and the other one is the toroidal magnetic field produced by a set of external coils. These fields maintain a time independent static Magnetohydrodynamical (MHD) equilibrium satisfying the equation \cite{hazeltine2003, freidberg1982}

\begin{equation}
    \mathbf{J} \times \mathbf{B} - \nabla p = \mathbf{0},
    \label{mhd_conservacao_momento}
\end{equation}
where $\mathbf{J}$, $\mathbf{B}$ and $p$ are, respectively, the plasma current density, the magnetic field, and the plasma pressure. 

Taking the internal product of Eq.~(\ref{mhd_conservacao_momento}) and $\mathbf{B}$ results in

\begin{equation}
    \mathbf{B} \cdot \nabla{p} = 0.
\end{equation}
Therefore, in confined plasmas, the magnetic field lines lie on isobaric magnetic surfaces \cite{freidberg1982}.

Magnetic surfaces can be of two types, rational and irrational according to their winding numbers. In the first case, i.e. rational surfaces, the field lines close after a certain number of toroidal and poloidal turns. While on irrational surfaces, the magnetic field lines never close and cover the entire surface.

The equilibrium magnetic configuration $\mathbf{B}_{E}$, 
which forms the integrable part, is the superposition of the poloidal $\mathbf{B}_\theta$ and toroidal $\mathbf{B}_\varphi$ magnetic fields

\begin{equation}
    \mathbf{B}_{E} = \mathbf{B}_{\theta} + \mathbf{B}_{\varphi}.
    \label{equilibrium_field}
\end{equation}

We analyze the plasma confinement for a tokamak of large aspect ratio $\epsilon = R/a \gg 1$, where $R$ is the tokamak major radius and $a$ is the plasma column radius. To localize a magnetic field line, we use pseudo-toroidal coordinates $(r, \theta, \varphi)$, where $(r, \theta)$ (analogous to the poloidal coordinate system) lie on a surface defined by a cross-section of the tokamak chamber located at a certain toroidal angle $\varphi$. Since we work with the condition $\epsilon \gg 1$, the toroidal vessel tokamak may be seen as a periodic cylinder, so that any length along it is given by $Z = R \varphi$.

The toroidal magnetic field, as a first approximation, will be taken as homogeneous. Therefore $\mathbf{B_{\varphi}} \approx B_0 \hat{z}$, where $\hat{z}$ is the unit vector parallel to the cylinder length. All the effects due to the toroidal geometry of the tokamak are considered \textit{a posteriori}, as a correction in the cylindrical equilibrium map described in the next section. Additionally, our poloidal magnetic field depends only on the radial coordinate, $\mathbf{B_{\theta}}= B_{\theta}(r)\hat{\theta}$, and is determined by the parabolic plasma current \cite{caldas1996}

\begin{equation}
    \mathbf{j}(r) = j_0 \left[ 1 - \left( \frac{r}{a} \right)^2\right]^{\gamma} \Theta(a - r) \hat{z},
    \label{densidade_de_corrente}
\end{equation}
where $\Theta$ is the Heaviside step function, so $\Theta = 1$ for $r \leq a$ and $\Theta = 0$ for $r > a$, $j_0$ is a constant, and $\gamma$ is an exponential factor to fit the experimental plasma current profiles. 

Since the magnetic field $\mathbf{B}_E$ is always tangent to the field line, we have 

\begin{equation}
    \mathbf{B}_E \times d \mathbf{l} = \mathbf{0},
    \label{equacao_linha_campo}
\end{equation}
where $d\mathbf{l}$ is an infinitesimal element of the magnetic field line. From this equation we determine, in the cylindrical approximation, a relationship between the toroidal and poloidal displacements of the magnetic field line and the equilibrium magnetic field components

\begin{equation}
    B_\theta dz - B_0 r d\theta = 0,
    \label{comp_field_line_eq}
\end{equation}
where $dz = R d\varphi$. As mentioned before, the effects of the toroidal geometry are considered \textit{a posteriori}.

From Eq.~(\ref{comp_field_line_eq}) we obtain the safety factor in the cylindrical approach

\begin{equation}
    q(r) \equiv \frac{1}{2 \pi} \int \nu d\theta= \frac{r B_0}{R B_{\theta}(r)}.
    \label{cilindrical_safety_factor}
\end{equation}
It measures the average slope of the local magnetic field line slope $\nu = \frac{d \varphi}{d \theta}$. The poloidal magnetic field magnitude $B_\theta(r)$ is determined by the Ampère's law for a current density given, in our case, by Eq.~(\ref{densidade_de_corrente}). 

The safety factor is an important parameter in the operation of tokamaks, with a direct correspondence with the field line winding number $\iota$ widely used in dynamical system analyses:

\begin{equation}
    \iota = 1 / q.
    \label{winding_number}
\end{equation}

Since the considered plasma current density is a monotonic function, the equilibrium field lines have a monotonic winding number profile. Consequently, in this case the field line equation describes a twist dynamical system commonly identified in the Hamiltonian description of plasma physics \cite{lichtenberg1992, evans2008}. Thus, our magnetic configuration can be described by a twist Hamiltonian system for which scenarios of bifurcations and onset of chaos are known \cite{hazeltine2003}.

In this article, the perturbation on the equilibrium system is given by the ergodic magnetic limiter. The magnetic limiter is composed of one or more rings of length $g << 2 \pi R$. 
The resulting magnetic field $\mathbf{B}$ is the superposition of the equilibrium magnetic field $\mathbf{B}_E$ and the magnetic field $\mathbf{B}_L$ produced by the ergodic limiter: 
\begin{equation}
    \mathbf{B} = \mathbf{B}_E + \mathbf{B}_L.
\end{equation}

Each ring that makes up the limiter has $m$ pairs of wires in the toroidal direction, and the electric current $I_h$ passing through the ring generates a magnetic field (for simplicity, we also represent the magnetic field of each ring as $\mathbf{B}_L$) with the following components \cite{caldas1996}:

\begin{equation}
    (B_L)_r = - \frac{\mu_0 g m I_h}{\pi b} \left( \frac{r}{b}\right)^{m-1} \sin{(m \theta)}
    \sum_{k = - \infty}^{\infty} \delta (\varphi - 2 \pi k),
\end{equation}

\begin{equation}
    (B_L)_\theta = - \frac{\mu_0 g m I_h}{\pi b} \left( \frac{r}{b}\right)^{m - 1} \cos{(m \theta)} 
    \sum_{k = - \infty}^{\infty} \delta(\varphi - 2 \pi k),
\end{equation}
where $\mu_0$ is the vacuum magnetic permeability, $b$ is the tokamak minor radius and $\delta$ is the Dirac delta function. The toroidal component $(B_L)_\varphi$ is neglected because it generates a magnetic field much smaller than the equilibrium toroidal magnetic field $B_\varphi$ \cite{pires2005}. We also neglect the plasma response to the magnetic field created by the limiter because, in our case, the perturbation amplitude is high only near the plasma edge, where the density is smaller than in the plasma core.

When the equilibrium system is perturbed, the equilibrium rational surface resonant with the perturbation gives rise to surrounding islands whose size depends on the perturbation strength. For strong enough perturbations, the irrational surfaces near the islands are destroyed and give rise to chaos, whereas farther irrational surfaces survive but are distorted.

Therefore, the perturbed plasma presents a mixed dynamics formed by regions where the plasma is confined by magnetic surfaces and regions with chaotic magnetic field lines and magnetic islands. The explanation for the appearance of magnetic islands and chaos and the survival of other magnetic surfaces, from the point of view of Hamiltonian dynamics, is given by the Poincaré-Birkhoff and the Kolmogorov-Arnold-Moser (KAM) theorems \cite{berry1978,lichtenberg1992}.

The winding number defined by Eq.~(\ref{winding_number}) has a physical meaning for our system. It determines the poloidal variation $\Delta \theta$ that a field line undergoes upon completing a toroidal turn on a magnetic surface as considered from a Poincaré surface of section $\varphi = 0$. Thus, we use the following expression whenever necessary to numerically determine $\iota$ 

\begin{equation}
    \iota = \lim_{k \to \infty} \frac{\sum_{l = 0}^k (\theta_{l+1} - \theta_l)}{2 \pi k}.
    \label{numerical_rotation_number}
\end{equation}
This expression gives the average over the poloidal variation $(\theta_{l + 1} - \theta_l)$ of the field line for $k$ toroidal rotations, with $k \to \infty$.

\section{Magnetic Field Mapping}
\label{Bmapping}

We consider the intersections of magnetic field lines with a toroidal section $Z$ of fixed coordinates. Such intersections are represented by a Poincaré map expressed in terms of an analytical symplectic two-dimensional map that describes how the field lines advance along the toroidal direction of the tokamak. 

The map we use is adapted from the Ullmann map \cite{ullmann2000}, which simulates the toroidal evolution of magnetic field lines along a tokamak with an ergodic magnetic limiter constituted of a single ring. In this article, we work with $N$ limiter rings equally spaced along the toroidal vessel, forming $N/2$ pairs. Each pair is responsible for creating a perturbative mode $(m, n)$, and each ring is spaced by $\Delta \varphi = \pi$ radians from its pair.

The new map, like the original one, is composed of two parts. The first part determines the spatial evolution of the magnetic field lines between two successive ring limiters, thus considering only the equilibrium field $\mathbf{B}_E$. The second part describes the local action of the limiter as an impulsive perturbation $\mathbf{B}_L$. Furthermore, the map is symplectic, i.e area preserving, because the field lines trajectories can be described by a Hamiltonian system \cite{boozer2005}.

Integrating the equilibrium Eq.~(\ref{comp_field_line_eq}) along $2 \pi / N$ radians in the toroidal direction, we find a relation between $\theta = \theta (\varphi')$, the poloidal position of the magnetic field line at the exit of the ring located at $\varphi'$, and $\theta^* = \theta(\varphi' + 2 \pi / N)$, the poloidal position of the magnetic field line at the entrance of the next ring at $\varphi' + 2 \pi / N$ 

\begin{equation}
    \int_{\varphi'}^{\varphi' + 2 \pi / N} d \varphi = \int_{\theta(\varphi')}^{\theta(\varphi' + 2\pi / N)}
    \frac{B_0 r}{B_\theta R} d \theta.
    \label{equilibrio_coordenada_poloidal}
\end{equation}

The magnetic field lines are radially constant in the cylindrical approach, so

\begin{equation}
    r = r^*.
    \label{equilibrio_coordenada_radial}
\end{equation}

One canonical transformation that results in the cylindrical map, Eqs.~(\ref{equilibrio_coordenada_poloidal}) and (\ref{equilibrio_coordenada_radial}), is

\begin{equation}
    G_{cil.}(r^*, \theta) = \theta r^* + \frac{2 \pi}{N} \int_0^{r^*} \frac{d r}{q(r)}.
    \label{geratriz_cilindrica}
\end{equation}
Therefore, the map given by Eqs.~(\ref{equilibrio_coordenada_poloidal}) and (\ref{equilibrio_coordenada_radial}) can be derived from the relations

\begin{equation}
    r = \frac{\partial G_{cil.}}{\partial \theta},
    \label{relacao_radial}
\end{equation}

\begin{equation}
    \theta^* = \frac{\partial G_{cil.}}{\partial r^*}.
    \label{relacao_poloidal}
\end{equation}
\begin{center}
\end{center}

However, since we want to consider the effects of the toroidal geometry of the tokamak, we adopt a toroidal correction similar to that given by Ullmann \cite{ullmann2000}. We assume that the toroidal-corrected generating function $G_{tor}(r^*, \theta)$ is given by the sum of Eq.~(\ref{geratriz_cilindrica}) and a toroidal correction, resulting in
\begin{equation}
    G_{tor.}(r^*, \theta) = G_{cil.}(r^*, \theta) + \frac{1}{N} \sum_{l = 1}^{\infty} \bar{a}_l \left ( \frac{r^*}{R}\right)^{l} \cos^{l} \theta,
    \label{geratriz_toroidal}
\end{equation}
where the coefficients $\bar{a_l}$ of summation are chosen to reproduce safety profiles of the equilibrium mapping observed in tokamaks \cite{ullmann2000}. 

Using again the relations between the new and old variables, Eqs.~(\ref{relacao_radial}) and (\ref{relacao_poloidal}), we obtain the equilibrium map with toroidal correction:

\begin{equation}
    r^* = \frac{r}{1 - \frac{1}{N} a_1 \sin{\theta}},
    \label{radial_equilibrio}
\end{equation}

\begin{equation}
    \theta^* = \theta + \frac{1}{N} \frac{2 \pi}{q(r^*)} +
    \frac{1}{N} a_1 \cos{\theta},
    \label{poloidal_equilibrio}
\end{equation}
where only the first term, $a_1 = \bar{a}_1/R$, of the summation in (\ref{geratriz_toroidal}) is considered because $(r / R)^l$ is negligible as $l$ grows. Given the magnetic field line position $(r, \theta)$ at the exit of one ring limiter, Eqs.~(\ref{radial_equilibrio}) and (\ref{poloidal_equilibrio}) determine the position $(r^*, \theta^*)$ of the field line at the entrance of the next limiter. 

The second part of the map is responsible for describing the perturbation on the equilibrium magnetic field line every time it passes through a limiter ring. To obtain this second part, we turn to the generating function perturbation $G_{pert.}$ of the Ullmann map \cite{ullmann2000}. However, the limiter rings must now work in pairs in order to produce a $(m, n)$ resonant perturbation mode. It means there must be a poloidal rotation between them, and thus we have

\begin{eqnarray}
    G_{pert.}(r, \theta^*) = \theta^* r - \frac{bC_i}{m_i - 1}
    \left (\frac{r}{b} \right)^{m_i - 1} \nonumber \\ \cos{[m_i(\theta^* + (k - 1) \alpha_i)]},
    \label{geratriz_limitador}
\end{eqnarray}
where $C_i = 2 \epsilon_i m_i g a^2 / q(a) R b^2$ is a dimensionless constant, $\epsilon_i = (I_h)_i / I_p$ is the ratio between the electric current $(I_h)_i$ in the limiter pair $i$ ($i = 1, 2, ..., N/2$) and the plasma current $I_p$, and $k = 1, 2$ is an index determining whether it is the first or second ring of the same pair $i$. We point out that the generating function perturbation Eq.~(\ref{geratriz_limitador}) for our case differs from the generating function in \cite{ullmann2000} only by the correction $(k - 1) \alpha_i$ in the argument of the cosine function. The factor $\alpha_i$ is the poloidal rotation between the limiters of the same pair

\begin{equation}
    \alpha_i \equiv \Delta \theta_i = \pi \frac{n_i}{m_i}.
    \label{giro_poloidal}
\end{equation}

The rationale behind Eq.~(\ref{giro_poloidal}) is the fact that the rings of the same pair need to have a poloidal rotation between them so that $m_i/n_i = \Delta \varphi' / \Delta \theta_i$, where $\Delta \varphi' = \pi$ is the toroidal variation between rings of the same pair.
This arrangement simulates pairs of wound wires, with helical pitch, such that the wires would close at every $m_i$ toroidal turns and $n_i$ poloidal turns, exciting a $(m_i, n_i)$ resonant perturbation mode. As a consequence, this procedure introduces the angle $\alpha_{i}$ containing the toroidal wave number $n_i$. This number and the poloidal wave number $m_{i}$ determine the winding number $\iota = n_i / m_i$ of the mode excited by a given limiter.

It is important to mention that due to the toroidal nature of the system, secondary poloidal modes $m_i \pm k$ ($k$ an integer and $k \neq 0$) may also be excited and modify the Poincaré section. Among these secondary modes, the most influencial are usually those with poloidal numbers $m_i  \pm 1$. However, for all the cases under study in this paper, the amplitudes of secondary modes are much smaller than the amplitude of the primary mode $m_i$. And since the secondary modes are resonant on their own $(m_i \pm k, n_i)$ surfaces, they do not interfere in the primary bifurcations of the main mode $(m_i, n_i)$ analyzed in the next sections.

 Applying to (\ref{geratriz_limitador}) a canonical transformation similar to Eqs.~(\ref{relacao_radial}) and (\ref{relacao_poloidal}), we obtain the relationship between the position $(r^*, \theta^*)$ of the magnetic field line immediately before passing through a limiter ring, and its position $(r, \theta)$ immediately after passing through this ring:

\begin{equation}
    r^* = r + \frac{bm_iC_i}{m_i - 1} \left( \frac{r}{b} \right)^{m_i - 1} \sin[m_i (\theta^* + (k - 1) \alpha_i)],
    \label{radial_perturbado}
\end{equation}

\begin{equation}
    \theta = \theta^* - C_i \left( \frac{r}{b} \right)^{m_i - 2}
    \cos[m_i(\theta^* + (k - 1) \alpha_i)].
    \label{poloidal_perturbado}
\end{equation}

The spatial sequence of a field line, in a toroidal turn, is determined by computing Eqs.~(\ref{radial_equilibrio}), (\ref{poloidal_equilibrio}), (\ref{radial_perturbado}) and (\ref{poloidal_perturbado}) $N$ times. In the next sections, we use the map formed by Eqs.~(\ref{radial_equilibrio}), (\ref{poloidal_equilibrio}), (\ref{radial_perturbado}) and (\ref{poloidal_perturbado}) to obtain two-dimensional Poincaré sections and winding number profiles, and we analyze different perturbed magnetic configurations. We consider limiter arrangements consisting of two pairs of rings that are able to simultaneously excite two primary modes, $(m_1, n_1) \neq (m_2, n_2)$, on the same rational invariant with winding number $\iota = n_1 / m_1 = n_2 / m_2$. Such arrangements are represented by $\{(m_1, n_1), (m_2, n_2)\}$, and both the toroidal $n_i$ and poloidal $m_i$ numbers are determined by the ergodic limiter.

\begin{table}[b]
\caption{\label{tab:TCABR}%
TCABR tokamak parameters \cite{silva2002}.
}
\begin{ruledtabular}
\begin{tabular}{lcr}
\textrm{Parameter}&
\textrm{Symbol}&
\textrm{Numerical value}\\ 
\colrule
major radios & $R$ & $0.61\,m$\\
minor radius & $b$ & $0.21\,m$\\
plasma column radius & $a$ & $0.18\,m$\\
equilibrium toroidal field & $B$ & $1.0\,T$\\
ergodic limiter length & $g$ & $0.08\,m$\\
\end{tabular}
\end{ruledtabular}
\end{table}

\section{Heteroclinic Island Bifurcations}
\label{Heteroclinic_Bifurcations}

The $(m, n)$ resonant perturbation mode creates a periodic trajectory which closes itself after $m$ toroidal and $n$ poloidal turns. According to the Poincaré-Birkhoff fixed point theorem \cite{birkhoff1927,berry1978,lichtenberg1992,voyatzis2002}, this trajectory is represented by $2sm$ periodic points, half stable and half unstable, with winding number $\iota = n/m$ on a two-dimensional Poincaré section. The nonzero positive integer $s$ is a topological index not determined by the theorem \cite{birkhoff1927,berry1978,lichtenberg1992}. In Refs.~\cite{Sousa_2013,sousa2014,sousa2015}, it is shown that $s$ corresponds to the number of isochronous island chains, i.e. island chains with the same winding number, present on the Poincaré sections. Furthermore, in this paper we observe that $s = 1$ for homoclinic island separatrixes, and $s > 1$ for heteroclinic separatrixes.

In this section, we present Poincaré surface of sections displaying the $s$ resonant island chains that appear around the $sm$ stable periodic points. We analyze different ergodic limiter arrangements, and observe that isochronous bifurcations are caused by increasing electric current, thus altering the number $s$ of chains. Whenever new island chains are created by isochronous bifurcations, their winding number and resonant surface are the same as the previous chains.

The computer simulations are performed with the parameters of the TCABR --- Tokamak Chauffage Alfvén Brésilien, located in the Plasma Physics Laboratory of the University of São Paulo \cite{pires2005}. Its main parameters are summarized in Table (\ref{tab:TCABR}).
Furthermore, we consider $a_1 = -0.02$ for the TCABR parameters, and we impose that the safety factor at the center of the plasma column is $q(r=0) = 1$. Thus, for each value of $q(a)$ chosen, the value of $\gamma$ in Eq.~(\ref{densidade_de_corrente}) is determined by the relation

\begin{equation}
    q(a) = (\gamma + 1) q(0).
\end{equation}

The Poincaré sections (for a fixed coordinate $Z$) represent the magnetic field lines in normalized rectangular coordinates $(X = \theta / 2 \pi, \, Y = (b - r) / b)$, which makes the visualization of the magnetic islands clear. We choose three different perturbative mode coupling scenarios to study. All scenarios are determined by an arrangement of the limiter pairs $\{(m_1, n_1), (m_2, n_2)\}$, with a given safety factor $q(a)$, keeping constant the electric current $(I_h)_1$ in the limiter component $(m_1, n_1)$, and varying the electric current $(I_h)_2$ in the other component $(m_2, n_2)$. For this reason, we use as a control parameter the ratio $\epsilon_i = I_i/I_p$ (with $I_i \equiv (I_h)_i$ for simplicity) between the electric current in the limiter pairs and the plasma current.

We perturb the invariant located at $r = a$ ($Y \approx 0.14$), the boundary between the plasma column and the vacuum. Moreover, the ratios $m_1 / n_1$ and $m_2/n_2$ are such that $m_1 / n_1 = m_2 / n_2 = q(a)$ with $(m_1, n_1) \neq (m_2, n_2)$. This causes the main perturbation in the plasma, due to the ergodic magnetic limiter, to be resonant at the surface with $q = q(a)$.

\subsection{Arrangement $\mathbf{\{(4, 1), (8, 2)\}}$}

\begin{figure*}[!tb]
\includegraphics[width=1.0\linewidth]{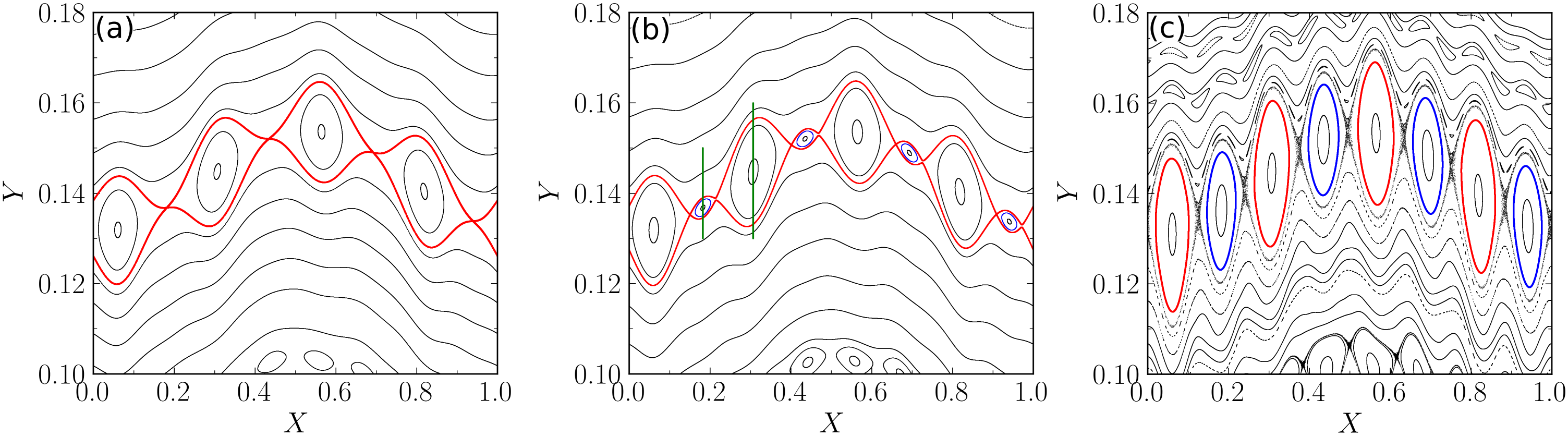}
    \caption{Arrangement $\{(4, 1), \,(8, 2)\}$ for $\epsilon_1 = 1.506 \times 10^{-3}$ and a)~$\epsilon_2 = 7.53 \times 10^{-4}$, b)~$\epsilon_2 = 1.205 \times 10^{-3}$, and c)~$\epsilon_2 = 2.259 \times 10^{-3}$. 
    }
    \label{arranjo_4_1_8_2}
\end{figure*}

\begin{figure}[!tb]
    \includegraphics[width=1.0\linewidth]{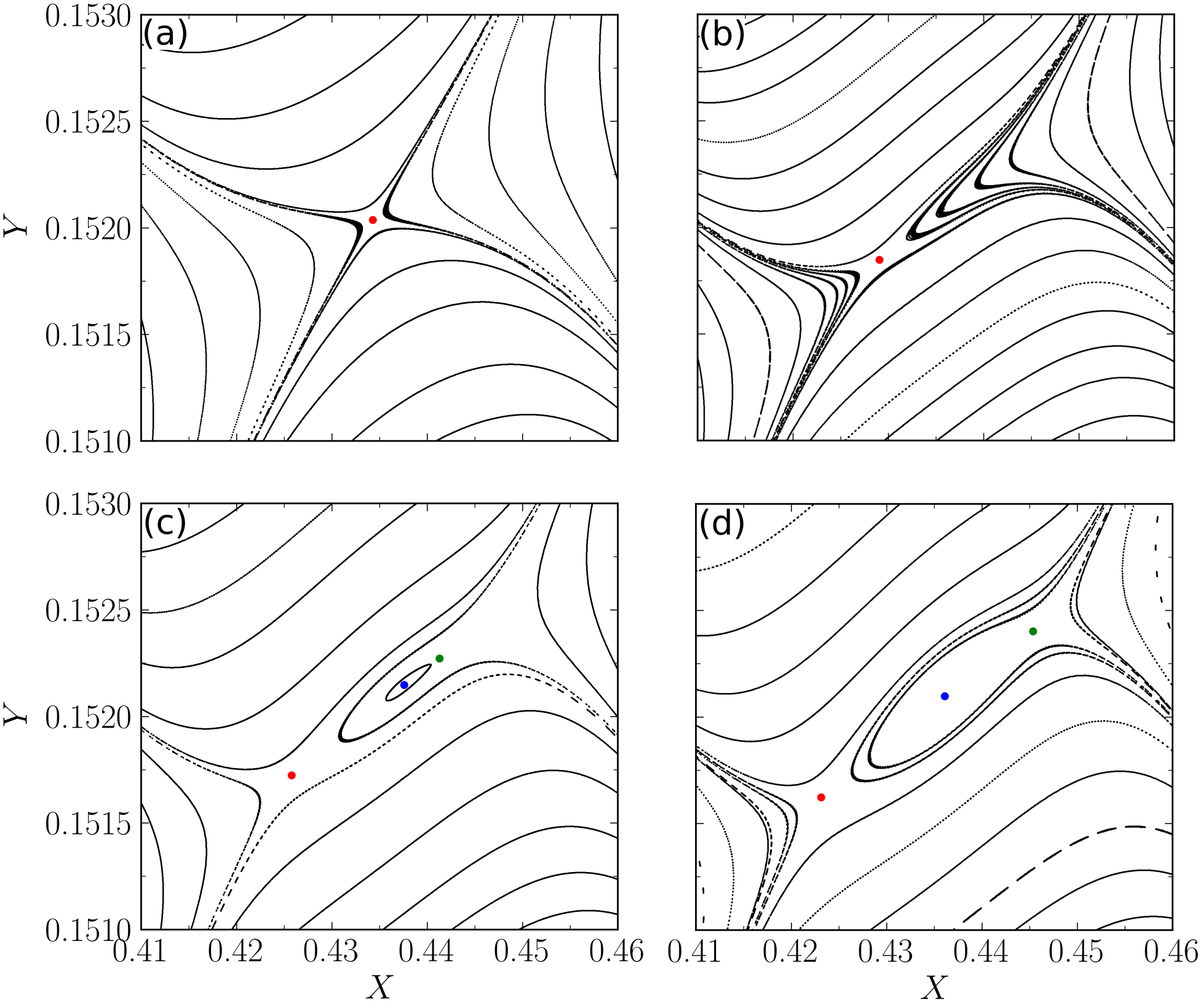}
    \caption{Arrangement $\{(4, 1), \,(8, 2)\}$ for $\epsilon_1 = 1.506 \times 10^{-3}$ and a)~$\epsilon_2 = 7.53 \times 10^{-4}$, b)~$\epsilon_2 = 8.28 \times 10^{-4}$, c)~$\epsilon_2 = 8.43 \times 10^{-4}$, and d)~$\epsilon_2 = 8.59 \times 10^{-4}$. In the transition from panels~(a) to (d), the hyperbolic red point moves to the left, and a saddle-node bifurcation occurs inside the separatrix of the initial $[4, 1]$ island chain.}
    \label{zoom_in_bifurcation_arrangement_4_1_8_2}
\end{figure}

\begin{figure}[!tb]
\includegraphics[width=0.7\linewidth]{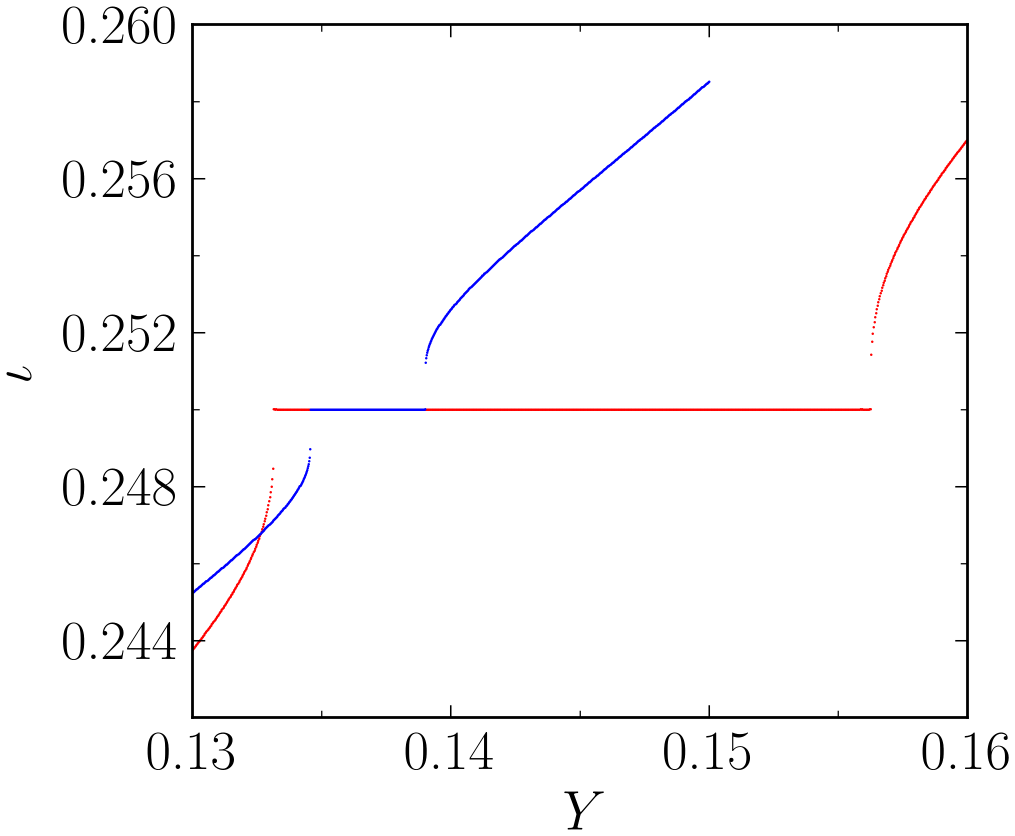}
    \caption{Winding number profile for the two green lines passing through both chains in Fig.~(\ref{arranjo_4_1_8_2}b). The blue curve is associated with the blue chain and the red curve with the red chain.
    }
    \label{winding_number_4_1_8_2}
\end{figure}

In this section, we consider the limiter arrangement which simultaneously excites modes $(4, 1)$ and $(8, 2)$ on the invariant line with $q(a) = 4$. For this arrangement, we observe the island chain bifurcation process depicted in Fig.~(\ref{arranjo_4_1_8_2}).
Initially, for values of $\epsilon_2$ much smaller than $\epsilon_1$, there is a single homogeneous chain excited by $\epsilon_1$. This chain represents the $(4, 1)$ mode, and it is formed by four islands as shown in Fig.~(\ref{arranjo_4_1_8_2}a) for $\epsilon_1 = 1.506 \times 10^{-3}$ and $\epsilon_2 = 7.53 \times 10^{-4}$.
The red trajectory in this figure highlights the approximated island separatrix. 
When the map is iterated, a point on this red curve, or on its internal islands, will jump from one island to its neighbor on the right. So, it is sufficient to have only one initial condition on one island to identify all the islands for this island chain. 

Increasing $\epsilon_2$ to a value close to $\epsilon_1$, each hyperbolic point of the $[4, 1]$ chain in Fig.~(\ref{arranjo_4_1_8_2}a) moves to the left, and a second island chain appears, as can be seen in Fig.~(\ref{arranjo_4_1_8_2}b). It means that, inside each island of the initial configuration in red in Fig.~(\ref{arranjo_4_1_8_2}a), a pair of elliptic-hyperbolic, i.e. stable-unstable, periodic points appears through a saddle-node bifurcation, as shown in details in Fig.~(\ref{zoom_in_bifurcation_arrangement_4_1_8_2}). The new elliptic points are the blue islands' center in Fig.~(\ref{arranjo_4_1_8_2}b).

\begin{figure*}[!bt]
    \includegraphics[width=1.0\linewidth]{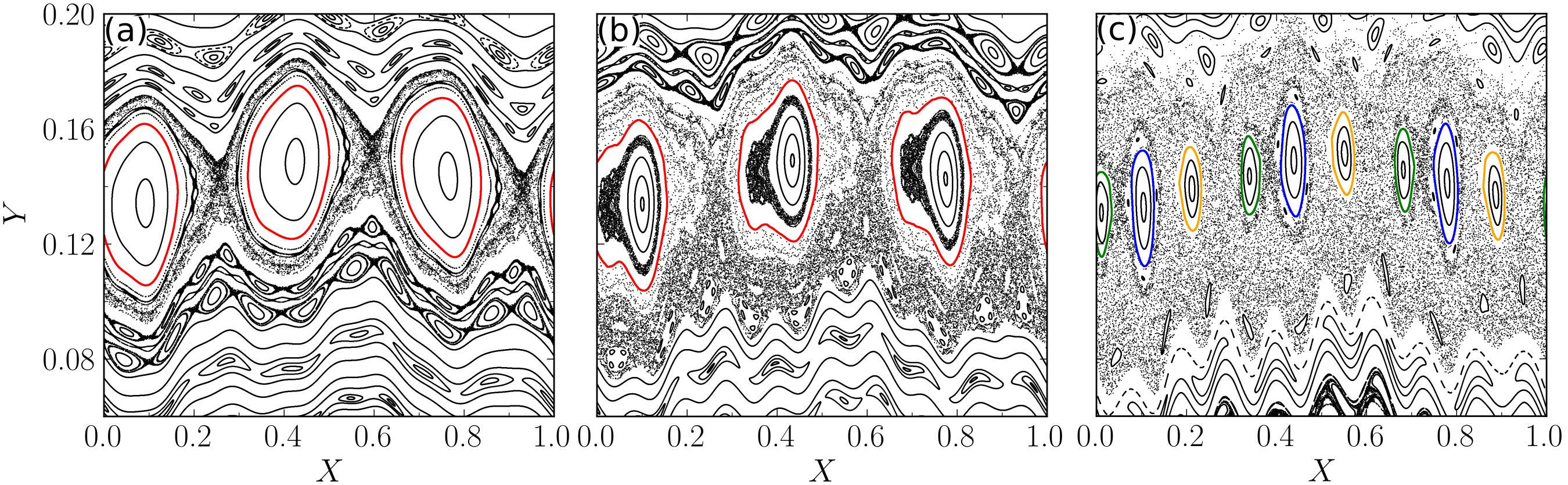}
    \caption{Arrangement $\{(3, 1), \,(9, 3)\}$ for $\epsilon_1 = 1.1296 \times 10^{-2}$ and a)~$\epsilon_2 = 3.389 \times 10^{-3}$, b)~$\epsilon_2 = 1.0167 \times 10^{-2}$, and c)~$\epsilon_2 = 2.2593 \times 10^{-2}$. Each initial island of the $(3, 1)$ mode in panel~(a) gives rise to three new islands of mode $(9, 3)$ in panel~(c).}
   \label{arranjo_3_1_9_3}
\end{figure*}

\begin{figure*}[!bt]
    \includegraphics[width=1.0\linewidth] {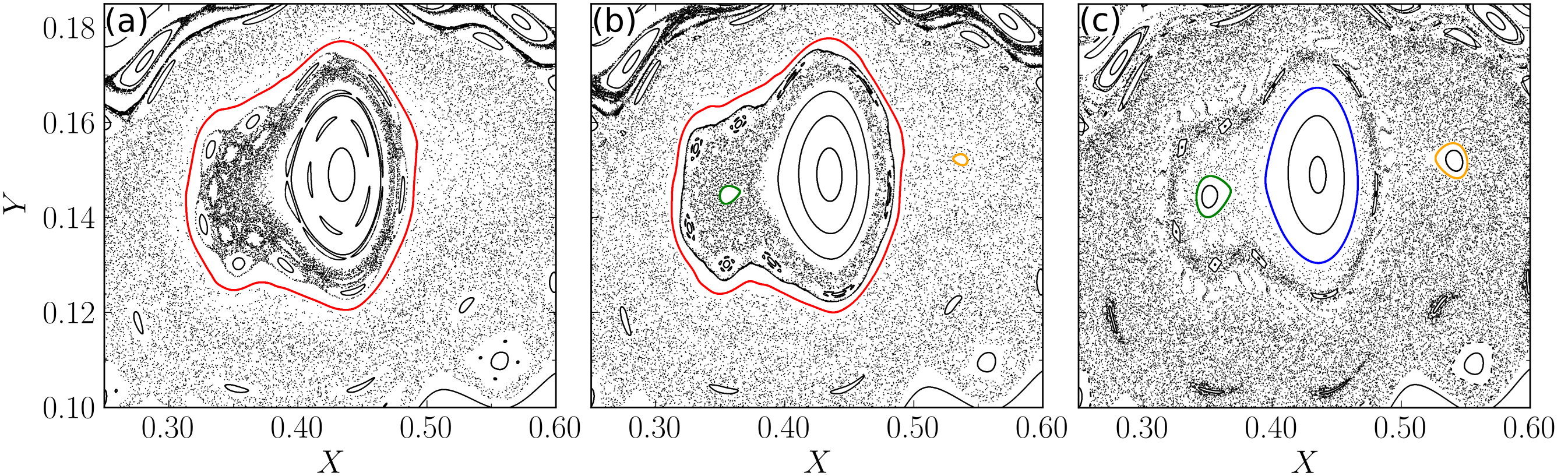}
    \caption{Bifurcation process for the arrangement $\{(3, 1),\,(9, 3)\}$: a)~$\epsilon_2 = 1.0167 \times 10^{-2}$, b)~$\epsilon_2 = 1.1296 \times 10^{-2}$, and c)~$\epsilon_2 = 1.2426 \times 10^{-2}$. For all cases $\epsilon_1 = 1.1296 \times 10^{-2}$. In panel~(b), the arrangement $\{(3, 1),\,(9, 3)\}$ is able to produce a mixture of the intermediate mode $(6, 2)$, in red and yellow, and the $(9, 3)$ mode, in green, black and yellow.}
    \label{ampliacao_3_1_9_3}
\end{figure*}

Figure (\ref{zoom_in_bifurcation_arrangement_4_1_8_2}) contains a sequence of mappings for increasing $\epsilon_2$ showing the onset of a pair of periodic points characterizing the saddle-node bifurcation that occurs inside each island of the initial $[4, 1]$ chain.
In Figs.~(\ref{zoom_in_bifurcation_arrangement_4_1_8_2}a) and (\ref{zoom_in_bifurcation_arrangement_4_1_8_2}b), there is only one red hyperbolic periodic point. However, increasing further the control parameter, we find in Fig.~(\ref{zoom_in_bifurcation_arrangement_4_1_8_2}c)  the original hyperbolic point in red and the new pair of elliptic and hyperbolic points, in blue and green respectively, created by the saddle-node bifurcation. As expected in this type of bifurcation, the distance between the two new points increases with the control parameter $\epsilon_2$, as we can see by comparing Figs.~(\ref{zoom_in_bifurcation_arrangement_4_1_8_2}c) and (\ref{zoom_in_bifurcation_arrangement_4_1_8_2}d).

When the saddle-node bifurcation occurs, the magnetic configuration changes from the homoclinic $[4, 1]$ chain in Fig.~(\ref{arranjo_4_1_8_2}a), where only one initial condition generates all four islands, to the heteroclinic $[8, 2]$ chains in Fig.~(\ref{arranjo_4_1_8_2}b). In this new situation, initial conditions iterated on a curve of the first chain jump to the second island on the right. Therefore, two sets of initial conditions are needed to produce all the islands of the two $[8, 2]$ chains, one for the islands in red, and another one for the blue islands. 

However, the appearance of the new blue chain in Fig.~(\ref{arranjo_4_1_8_2}b) occurs inside the red separatrix of the initial chain. If we keep increasing the parameter $\epsilon_2$, the separatrix breaks up. This scenario is shown in Fig.~(\ref{arranjo_4_1_8_2}c), where there is no longer a separatrix wrapping both $[8, 2]$ chains as in Fig.~(\ref{arranjo_4_1_8_2}b). Instead of a separatrix, there is a layer of chaos due to the increment of $\epsilon_2$ to $2.259 \times 10^{-3}$. It is also evident the increasing size of the blue island chain for $\epsilon_2 > \epsilon_1$. 

To show that both island chains appearing in Fig.~(\ref{arranjo_4_1_8_2}b) have the same $\iota$, and for this reason are isochronous \cite{egydio1992,voyatzis2002,Sousa_2013,sousa2014,sousa2015}, we calculate the winding number profile with the use of expression (\ref{numerical_rotation_number}). The lines in green in Fig.~(\ref{arranjo_4_1_8_2}b) represent two sets of initial conditions and the profiles formed by these sets are shown in Fig.~(\ref{winding_number_4_1_8_2}). Each winding number profile has the same color of the island it passes through in Fig.~(\ref{arranjo_4_1_8_2}b). The plateaus in Fig.~(\ref{winding_number_4_1_8_2}) correspond to the winding number of the islands, and both are equal to $\iota = 1/4 = 0.25$.

\subsection{Arrangement $\mathbf{\{(3, 1), (9, 3)\}}$}
\label{Arragement_(3,1)_(9,3)}

The perturbation now occurs on the surface $q(a) = 3$ with the arrangement $\{(3, 1), (9, 3)\}$. We fix the electric current in the limiter pair responsible for producing the perturbative mode $(3, 1)$, such that $\epsilon_1 = 1.1296 \times 10^{-2}$. In Fig.~(\ref{arranjo_3_1_9_3}), we show  Poincaré sections for different values of $\epsilon_2$ that generates the $(9, 3)$ mode. The values of $\epsilon_1$ and $\epsilon_2$ are chosen according to expected experimental values. However, in Fig.~(\ref{arranjo_3_1_9_3}) all the resonant islands are separated by chaotic areas. In the Appendix, we consider lower values of the limiter currents, and present Poincaré sections in which the separatrixes can be identified, corroborating the results discussed in this section. 

In Fig.~(\ref{arranjo_3_1_9_3}a), we only see the islands of the $[3, 1]$ chain, despite the value $\epsilon_2 = 3.389 \times 10^{-3}$ for the $(9, 3)$ mode amplitude.
When we increase $\epsilon_2$ to $1.0167 \times 10^{-2}$ as in Fig.~(\ref{arranjo_3_1_9_3}b), there is no primary bifurcation yet. However, the initial $[3, 1]$ islands are quite deformed. These islands undergo a stretching (in the horizontal direction in the coordinate system we are using), and chaos is noticeable inside them. Fig.~(\ref{ampliacao_3_1_9_3}a) shows an amplification of a $[3, 1]$ island of Fig.~(\ref{arranjo_3_1_9_3}b). The chaos seen inside the islands of this primary chain is due to secondary bifurcations: chains of small secondary islands start to appear around the elliptical points of the $[3, 1]$ chain for values of $\epsilon_2$ between those used in Figs.~(\ref{arranjo_3_1_9_3}a) and (\ref{arranjo_3_1_9_3}b). As $\epsilon_2$ increases, the chaos associated with the hyperbolic points of these secondary islands spreads inside the $[3, 1]$ primary islands.

Increasing $\epsilon_2$ to $2.2593 \times 10^{-2}$, the initial $[3, 1]$ chain in Fig.~(\ref{arranjo_3_1_9_3}a) gives rise to the heteroclinic $[9, 3]$ chains in Fig.~(\ref{arranjo_3_1_9_3}c). Each of the three initial red islands in Fig.~(\ref{arranjo_3_1_9_3}a) splits into three new islands due to heteroclinic bifurcations. Three different sets of initial conditions in successive islands are needed to produce the $[9, 3]$ chains. 
In Fig.~(\ref{arranjo_3_1_9_3}c), the highlighted islands in green, blue, and yellow are created using three different initial conditions.

We examine in more detail how the bifurcation process occurs for this arrangement. In Fig.~(\ref{ampliacao_3_1_9_3}a), we reproduce the case $\epsilon_2 = 1.0167 \times 10^{-2}$, as in Fig.~(\ref{arranjo_3_1_9_3}b), for a single island since the other two islands bifurcate in the same manner. For this value of $\epsilon_2$, an isochronous primary bifurcation has not yet occurred, and the chaos seen inside the primary islands is due to secondary bifurcations as commented before.

Increasing $\epsilon_2$ such that $\epsilon_2 = \epsilon_1 = 1.1296 \times 10^{-2}$ as in Fig.~(\ref{ampliacao_3_1_9_3}b), in addition to the initial island in red, two other independent islands belonging to other isochronous chains arise. These two saddle-noddle bifurcations are independent and have different thresholds, with the green island appearing for lower values of $\epsilon_2$ than the yellow one, as discussed in the Appendix. The new island in green appears inside the initial island in red. However, the other new island in yellow appears outside the island in red. This scenario is a mixture of the $(6, 2)$ mode presenting two independent island chains in red and yellow, and the $(9, 3)$ mode with three chains in green, black and yellow.

For $\epsilon_2 = 1.2426 \times 10^{-2} > \epsilon_1$ as in Fig.~(\ref{ampliacao_3_1_9_3}c), the island in red is broken and only the $(9, 3)$ mode with its three chains are observed: one in green, one in blue, and another one in yellow. Two saddle-noddle bifurcations give rise to the islands in green and yellow seen in Figs.~(\ref{ampliacao_3_1_9_3}b) and (\ref{ampliacao_3_1_9_3}c). We point out that even though the red island is broken in Fig.~(\ref{ampliacao_3_1_9_3}c), it is replaced by small secondary islands forming a stickiness layer in which chaotic trajectories remain confined for a great number of iterations before escaping to the chaotic sea.

\subsection{Arrangement $\mathbf{\{(6, 2), (15, 5)\}}$}
\label{Section_(6,2)_(15,5)}

\begin{figure*}[!bt]
    \includegraphics[width=0.9\linewidth] {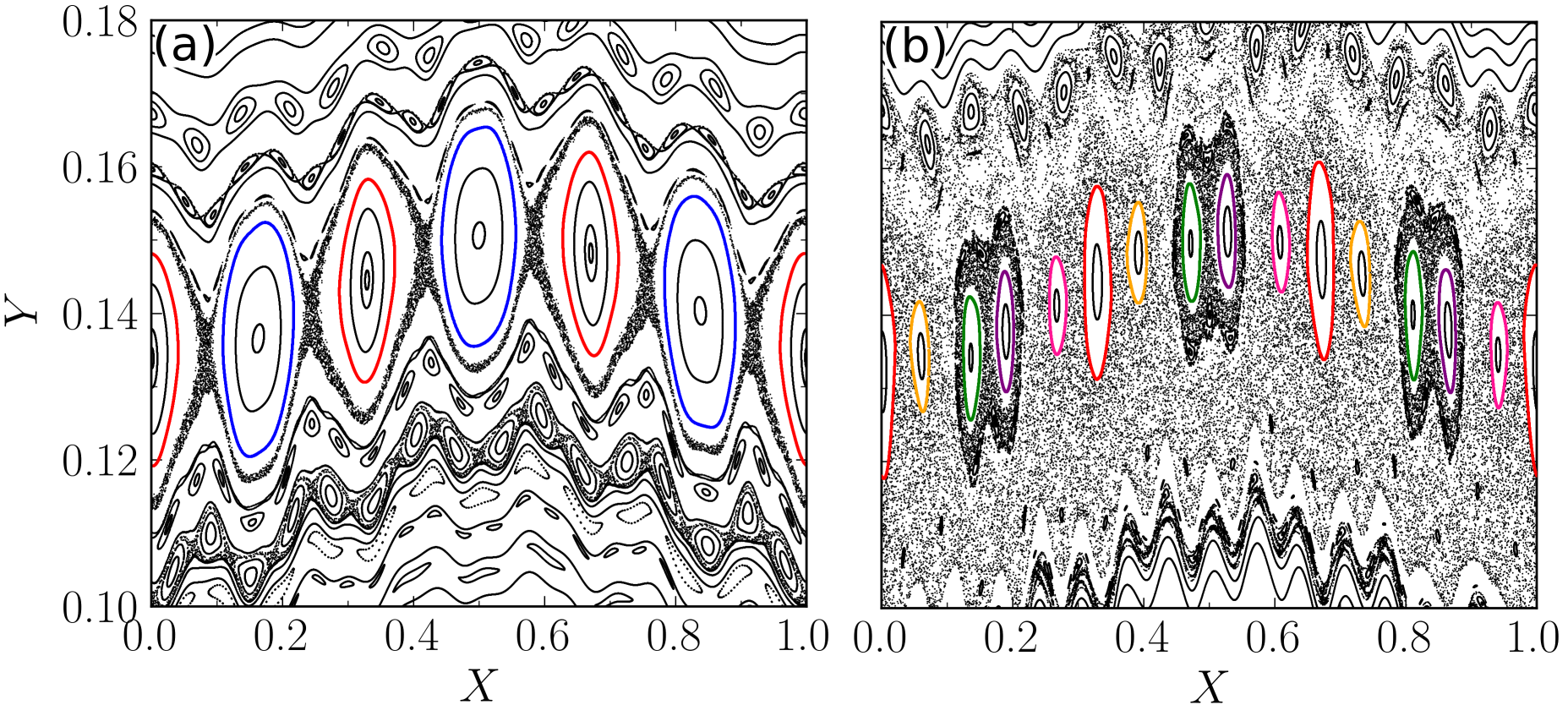}
    \caption{Arrangement $\{(6, 2), \,(15, 5)\}$ for $\epsilon_1 = 5.648 \times 10^{-3}$ and a)~$\epsilon_2 = 2.824 \times 10^{-3}$ and b)~$\epsilon_2 = 1.6944 \times 10^{-2}$. The isochronous islands that form the initial $[6, 2]$ configuration in panel~(a) bifurcate in two different ways, giving rise to the final $[15, 5]$ configuration in panel~(b).}
    \label{arranjo_6_2_15_5}
\end{figure*}

\begin{figure}[!bt]
    \centering
    \includegraphics[width=1.0\linewidth]{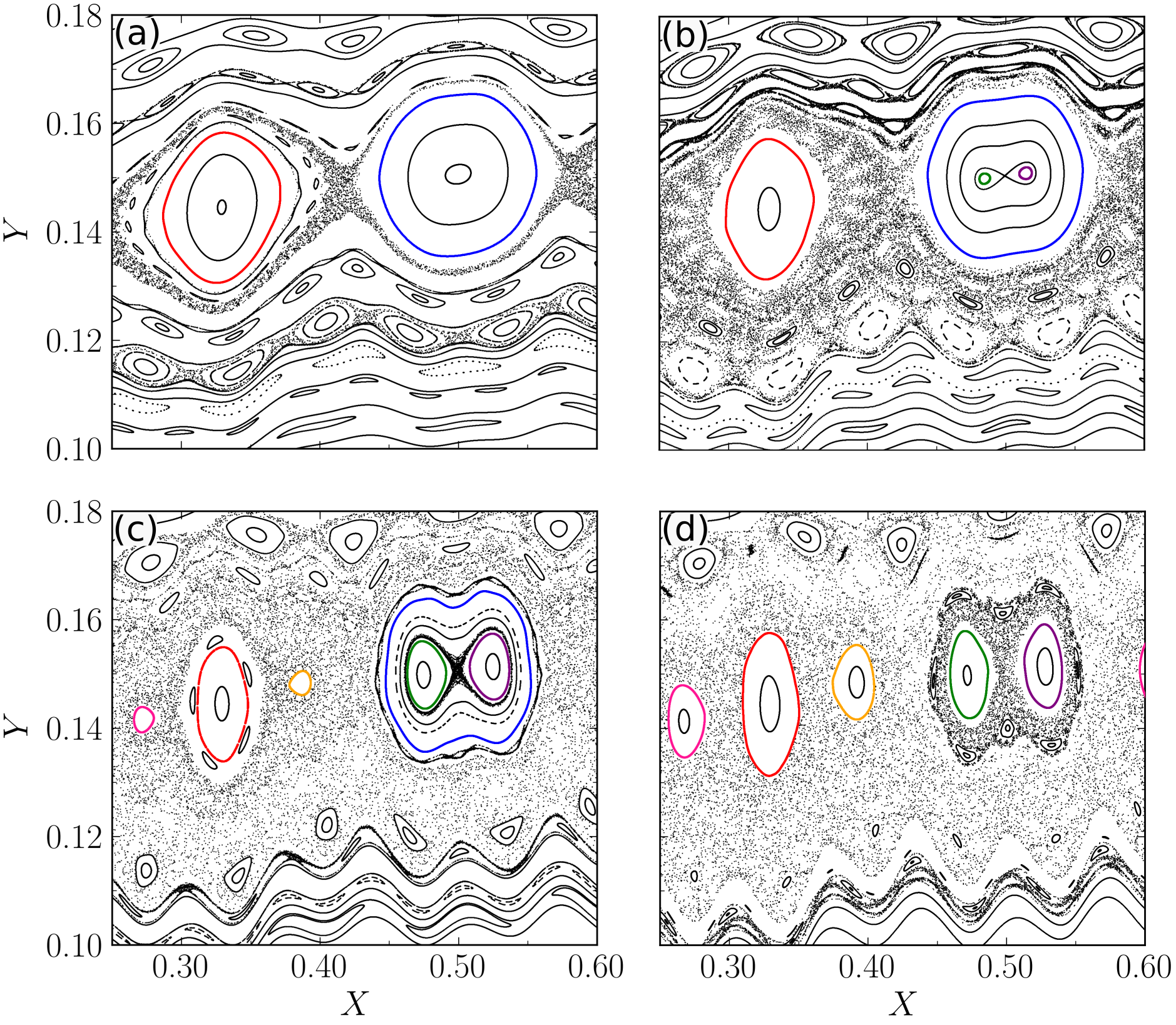}
    \caption{Zoom of arrangement $\{(6, 2), \,(15, 5)\}$: a)~$\epsilon_2 = 2.824 \times 10^{-3}$, b)~$\epsilon_2 = 5.648 \times 10^{-3}$, c)~$\epsilon_2 = 1.1296 \times 10^{-2}$, and d)~$\epsilon_2 = 1.6944 \times 10^{-2}$. For all cases $\epsilon_1 = 5.648 \times 10^{-3}$. As $\epsilon_2$ is increased, the islands in red and blue in panel~(a) undergo different bifurcations. In panel~(b), we observe a mixture of the $(6, 2)$ and $(9, 3)$ modes in [red, blue] and [red, green, purple] respectively. Panel~(c) exhibits a combination of the $(12, 4)$ and $(15, 5)$ modes in [pink, red, yellow, blue] and [pink, red, yellow, green, purple] respectively. Finally, panel~(d) displays the final $[15, 5]$ configuration.}
    \label{ampliacao_6_2_15_5}
\end{figure}

\begin{figure}[!bt]
    \centering
    \includegraphics[width=1.0\linewidth]{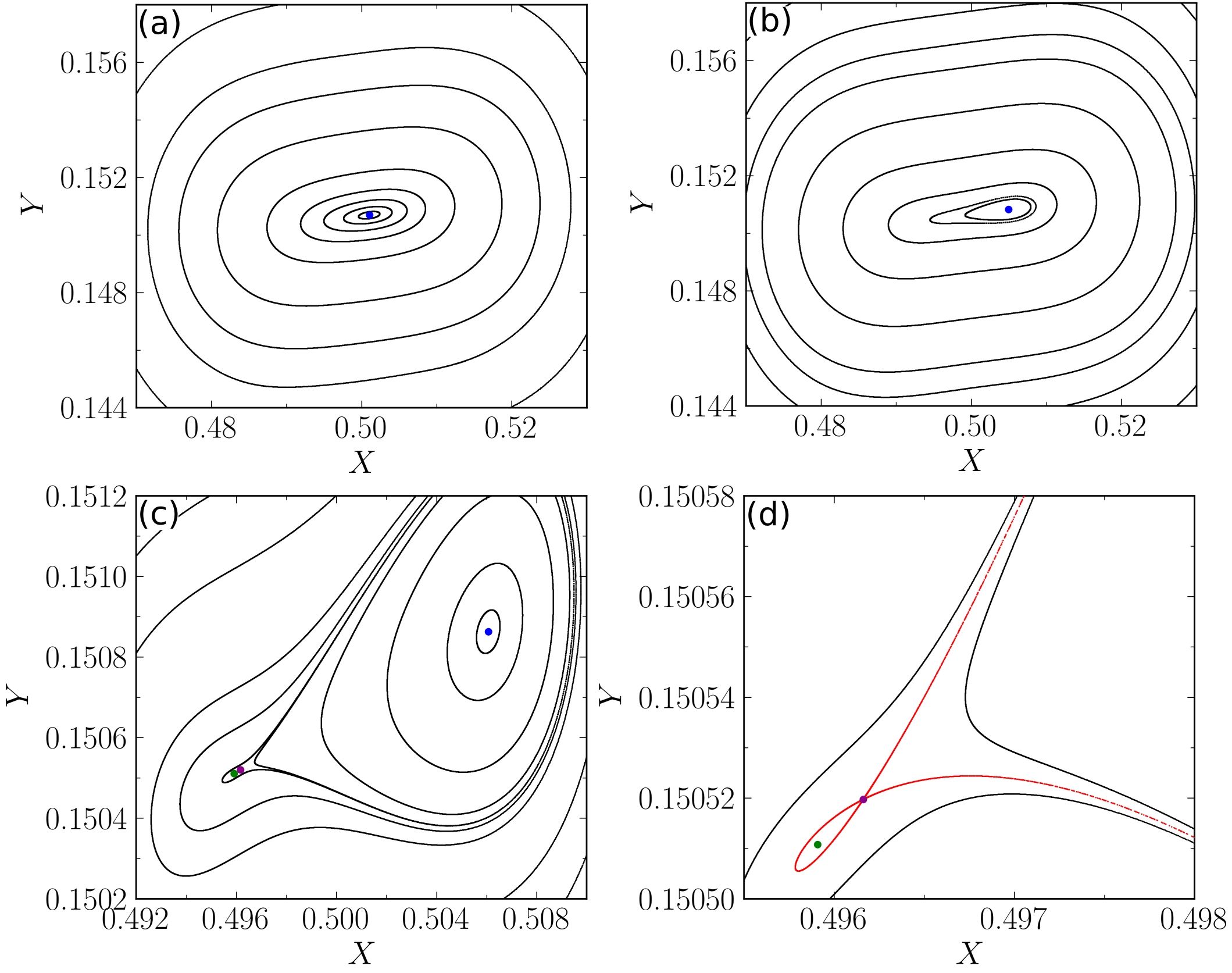}
    \caption{Arrangement $\{(6, 2), (15, 5)\}$ for $\epsilon_1 = 5.648 \times 10^{-3}$ and a)~$\epsilon_2 = 3.954 \times 10^{-3}$, b)~$\epsilon_2 = 4.293 \times 10^{-3}$, and c)~$\epsilon_2 = 4.377 \times 10^{-3}$. Panel~(d) shows an amplification of panel~(c). The elliptical blue point in (a) moves to the right as in panels~(b) and (c), and a saddle-node bifurcation happens as shown in panels~(c) and (d).}
    \label{zoom_in_bifurcation_6_2_15_5}
\end{figure}

The second case for $q(a) = 3$ that we analyze passes through more intermediate stages than the previous one. In Fig.~(\ref{arranjo_6_2_15_5}a), $\epsilon_1 = 5.648 \times 10^{-3} > \epsilon_2 = 2.824 \times 10^{-3}$, and the initial $[6, 2]$ chains are already heteroclinic, formed by two sets of islands in red and blue that do not present a common separatrix. This situation characterizes our initial configuration.

Meanwhile, for $\epsilon_2 = 1.6944 \times 10^{-2} \gg \epsilon_1$, Fig.~(\ref{arranjo_6_2_15_5}b) shows the final $[15, 5]$ configuration, i.e. a set of $5$ chains in the colors pink, red, yellow, green, and purple, each chain with three islands. This is the final stage and no further primary bifurcation happens as we keep increasing $\epsilon_2$.

Figure (\ref{ampliacao_6_2_15_5}) shows only two islands of the $[6, 2]$ initial chain, namely, the islands in red and blue located between $X = 0.25$ and $X = 0.60$. Each of these islands bifurcates in different ways, with the same bifurcation process occurring for all the other pairs of blue and red islands.

In Fig.~(\ref{ampliacao_6_2_15_5}a), $\epsilon_2 = 2.824 \times 10^{-3} < \epsilon_1 = 5.648 \times 10^{-3}$ and the bifurcations have not happened yet. But we can see different deformations for the two islands. The island in red assumes a shape similar to a rhombus, while the island in blue has an elliptical shape.

For $\epsilon_2 = 5.648 \times 10^{-3} = \epsilon_1 = 5.648 \times 10^{-3}$ as in Fig.~(\ref{ampliacao_6_2_15_5}b), a saddle-node bifurcation already occurred inside the blue island. Simultaneously with that, a mixed stage appears, where we can observe the initial $(6, 2)$ mode with its big red and blue islands, together with the $(9, 3)$ mode presenting a big island in red and two small islands in gree and purple. In the transition of Fig.~(\ref{ampliacao_6_2_15_5}a) to Fig.~(\ref{ampliacao_6_2_15_5}b), the original elliptical point of the blue island is displaced to the right, while a new pair of hyperbolic and elliptic points is created. This is a saddle-node bifurcation just like the one shown in Fig.~(\ref{zoom_in_bifurcation_arrangement_4_1_8_2}).

For $\epsilon_2 = 1.1296 \times 10^{-2} > \epsilon_1 = 5.648 \times 10^{-3}$ as in Fig.~(\ref{ampliacao_6_2_15_5}c), another bifurcation occurred. The small islands in pink and yellow appear outside the initial chain in red. We have thus another mixed stage formed by the $(12, 4)$ and $(15, 5)$ modes, which contain the islands in [pink, red, yellow, blue] and [pink, red, yellow, green, purple] respectively. It is interesting to note that the closed blue curves keep the green and purple islands isolated from the chaotic sea, whereas the pink, red and yellow islands are immersed in chaos.

Finally, for even higher values of $\epsilon_2$ as in Fig.~(\ref{ampliacao_6_2_15_5}d), the island in blue is broken but the two internal islands in green and purple remain. We thus have the final $[15, 5]$ configuration. In this final configuration, all the islands have comparable sizes and are immersed in the chaotic sea. However, the green and purple islands are surrounded by a stickiness layer which replaces the broken blue islands.

In Fig.~(\ref{ampliacao_6_2_15_5}c), the islands in pink and yellow arise through saddle-node bifurcations without any modification in the elliptic point of the red island. On the other hand, as mentioned before, a saddle-node bifurcation gives rise to the green and purple islands inside the island in blue, making the original elliptic point in the center move to the right before the saddle-node bifurcation takes place. We discuss now this process in detail.

Figure (\ref{zoom_in_bifurcation_6_2_15_5}) contains a sequence of mappings for increasing values of $\epsilon_2$, which shows the onset of the saddle-node bifurcation that occurs inside the blue island of the initial $[6, 2]$ chain. In Figs.~(\ref{zoom_in_bifurcation_6_2_15_5}a) and (\ref{zoom_in_bifurcation_6_2_15_5}b), there is only one elliptic periodic point. However, by increasing further the control parameter, we observe in Fig.~(\ref{zoom_in_bifurcation_6_2_15_5}c) the original elliptic point in blue, and the new pair of hyperbolic and elliptic points in brown and green, respectively, created by the saddle-node bifurcation. Fig.~(\ref{zoom_in_bifurcation_6_2_15_5}d) shows an amplified region around the new periodic points and their separatrix in red.

\section{Shearless Bifurcations}
\label{Shearless_Bifurcations}

\begin{figure}[!tb]
    \centering
    \includegraphics[width=1.0\linewidth]{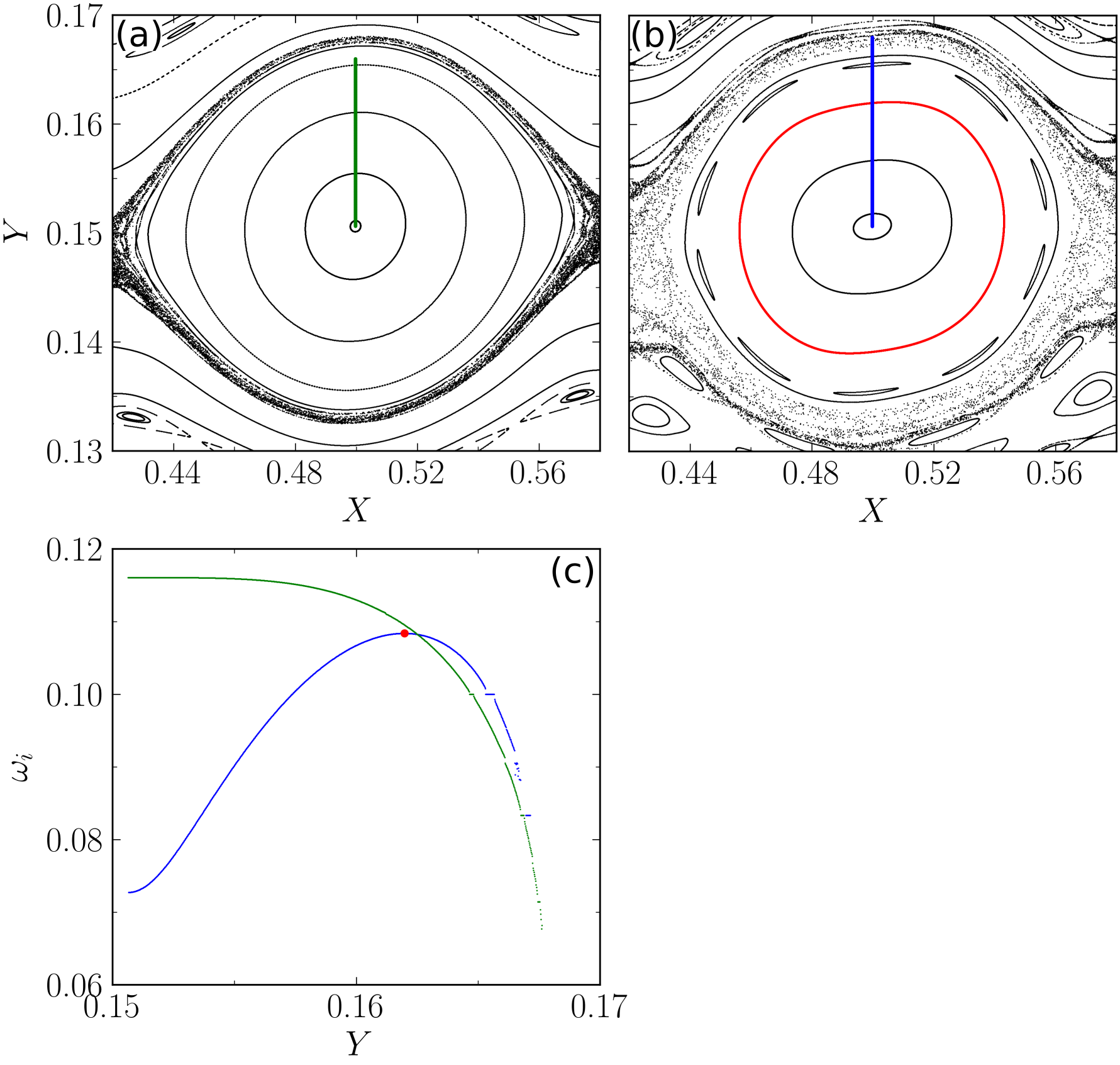}
    \caption{The type of internal winding number profile changes as $\epsilon_2$ increases. The island for $\epsilon_2 = 6.78 \times 10^{-4}$ in (a) has a monotonic internal winding number profile. But it changes to a nonmonotonic profile for $\epsilon_2 = 2.824\times10^{-3}$ in (b). Both profiles can be seen in (c). In all cases $\epsilon_1 = 5.648 \times 10^{-3}$.}
    \label{islands_and_winding_number}
\end{figure}

In the previous section, we analyzed the onset of isochronous primary islands created by the perturbing electric currents passing through the ergodic limiter. In this section, we present an example of secondary islands, known as twin islands \cite{morrison2000, del_castillo_2000}, created by the perturbations coupling. 

As mentioned in Section \ref{Bmapping}, we consider an equilibrium magnetic field whose field lines present monotonic winding number profiles, as can be seen in Fig.~(\ref{winding_number_4_1_8_2}). However, as reported in Ref.~\cite{abud2014}, the magnetic perturbation can modify these profiles such that, for a critical parameter, they become locally nonmonotonic, originating a local nontwist region with a shearless invariant inside primary islands. Increasing the control electric current above the critical value, two chains of secondary twin islands appear, one chain on each side of the shearless invariant, and both chains with the same winding number. This process corresponds to a shearless bifurcation \cite{morrison2000, del_castillo_2000, abud2012}.

We consider the arrangement $\{(6, 2)$, $(15, 5)\}$ of Section \ref{Section_(6,2)_(15,5)} in order to study the onset of a shearless bifurcation in our magnetic configuration. The islands under study belong to the blue chain in Figs.~(\ref{ampliacao_6_2_15_5}a)-(\ref{ampliacao_6_2_15_5}c). We determine the islands internal winding number profile $\omega_i$ for different values of $\epsilon_2$ and $X_0 = 0.5$.

\begin{figure}[!tb]
    \centering
    \includegraphics[width=1.0\linewidth]{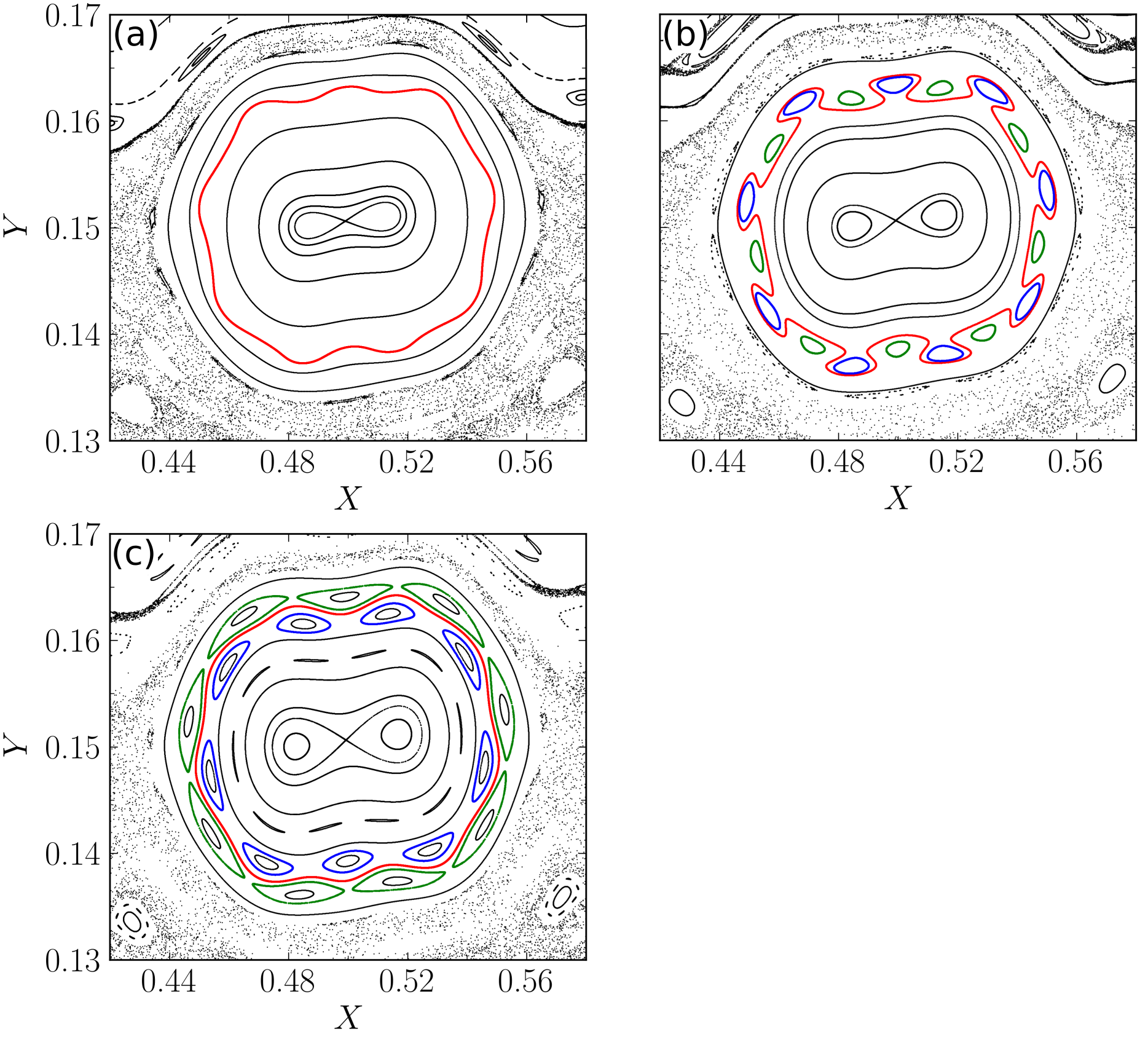}
    \caption{The perturbation parameters are $\epsilon_1 = 5.648 \times 10^{-3}$ and a)~$\epsilon_2 = 5.083\times 10^{-3}$, b)~$\epsilon_2 = 5.648\times 10^{-3}$, and c)~$\epsilon_2 = 6.213 \times 10^{-3}$. For panel~(a), the internal winding number profile is monotonic. However, it changes to nonmonotonic for panels~(b) and (c). Moreover, for these last two cases, it is possible to see the arising of secondary twin islands and their separatrix reconnection process.}
    \label{shearless_islands}
\end{figure}

The internal winding number formula is analogous to that given by expression (\ref{numerical_rotation_number}). But in that case, we look at the global poloidal variation of the field line with respect to a toroidal turn, whereas for the internal winding number, we determine the local poloidal variation with respect to the island center.

The internal winding number profile of an island is obtained by the numerical method described in Ref.~\cite{abud2012}. It consists in averaging on the island the local poloidal variation $P_{n}OP_{n + 1}$ of consecutive points, $P_{n}$ and $P_{n+1}$, for a number $n \to \infty$ of iterations such that 

\begin{equation}
    \omega_i = \lim_{n \to \infty} \frac{1}{2 \pi n}
    \sum_{n = 1}^{\infty} P_n O P_{n+1},
\end{equation}
where $P_{n}OP_{n+1}$ is the angle formed by the consecutive points when measured from the island center $O$.

The $\omega_i$ profiles for Figs.~(\ref{islands_and_winding_number}a) and (\ref{islands_and_winding_number}b) are shown in Fig.~(\ref{islands_and_winding_number}c). In both cases, we iterate $10^3$ initial conditions forming a line that starts in the island center (the elliptical point), and continues to the island edge. Each initial condition along this line is iterated for $n = 1.5 \times 10^5$. The points that converge are plotted on a $\omega_i \times Y$ graph. When the perturbation parameters are $\epsilon_1 = 5.648 \times 10^{-3}$ and $\epsilon_2 = 6.78 \times 10^{-4}$ as in Fig.~(\ref{islands_and_winding_number}a), the internal winding number profile is monotonic as shown by the green curve in Fig.~(\ref{islands_and_winding_number}c), where the winding number always decreases as the initial conditions move away from the island centre.

Increasing $\epsilon_2$ eventually makes the internal profile nonmonotonic. In Fig.~(\ref{islands_and_winding_number}b), $\epsilon_2 = 2.824 \times 10^{-3}$ and the island internal profile is shown in blue in Fig.~(\ref{islands_and_winding_number}c). This blue curve initially increases, reaching a maximum value highlighted as a red dot, and then it starts to decrease. The maximum value of $\omega_i$ corresponds to the shearless curve. The internal profile remains nonmonotonic until the appearance of the new green and purple islands of Figs.~(\ref{arranjo_6_2_15_5}) and (\ref{ampliacao_6_2_15_5}). The invariant red curve in Fig.~(\ref{islands_and_winding_number}c) is the shearless curve.

In resume, the monotonic internal profile becomes nonmonotonic, and the mapping changes from twist to nontwist inside the islands under study. The increase in the parameter $\epsilon_2$ (the perturbation amplitude of the $(15, 5)$ mode in the plasma column) alters the monotonic nature of the internal winding number profile of the islands that compose the $(6, 2)$ mode. We also notice that the bifurcation in the island center only takes place after the internal profile changes from monotonic to nonmonotonic.

In addition to the onset of the internal shearless invariant, we also analyze the emergence of secondary twin islands as $\epsilon_2$ increases. For $\epsilon_2 = 5.083 \times 10^{-3}$ as in Fig.~(\ref{shearless_islands}a), the central bifurcation already occurred, but not the arising of the two twin island chains. However, for $\epsilon_2 = 5.648 \times 10^{-3}$, it is possible to notice the secondary twin islands already formed inside the main island, as shown in Fig.~(\ref{shearless_islands}b). The green chain is on one side of the red torus, while the blue chain is on the other side. Finally, for $\epsilon_2 = 6.213 \times 10^{-3}$ as in Fig.~(\ref{shearless_islands}c), the separatrix reconnection process \cite{egydio1992, corso1998, wurm2004} took place, a feature commonly observed in nontwist systems, such that the twin green and blue chains exchange positions with respect to the red shearless curve.

\section{Conclusions}
\label{Conclusions}

We analyzed the magnetic field line dynamics of a plasma confined in a large aspect ratio tokamak and perturbed by an external ergodic magnetic limiter formed by two pairs of coils. Each limiter pair has its own electric current, and they generate magnetic perturbations with different wave numbers but the same winding number. Therefore, both perturbations are resonant on the same invariant surface, creating isochronous, i.e. with the same winding number, magnetic islands around that invariant. 

One pair of coils excites a mode with wave numbers $m_1$ and $n_1$, and the other pair a mode with wave numbers $m_2$ and $n_2$. When applied alone, each mode creates $m_1$ or $m_2$ isochronous islands in phase space. However, when the perturbations act simultaneously, the number of isochronous islands varies as a function of the perturbing limiter currents. 

We analyzed different $\{(m_1, n_1), (m_2, n_2)\}$ limiter arrangements by keeping constant the electric current $I_1$ in the limiter pair that generates the $(m_1, n_1)$ mode, and making $I_2$ vary in the pair producing the $(m_2, n_2)$ mode. When the currents are not on the same order of magnitude, the phase space displays the $m_i$ isochronous islands produced by the limiter pair with higher current. On the other hand, for currents with the same order of magnitude, we observed heteroclinic bifurcations creating new isochronous island chains as $I_2$ is increased.

For the considered ergodic limiter arrangements, we described isochronous bifurcations characterizing three ways in which the phase space may change from one island chain configuration to the other. One of them, such as the $\{(4, 1), (8, 2)\}$ arrangement, consists in the direct passage of the $(m_1, n_1)$ mode to the $(m_2, n_2)$ mode as $I_2$ increases. The other two types of transition occur through intermediate island chain configurations in which the phase space displays mixed states with different $(m_i, n_i)$ modes. In some cases, only one intermediate configuration is possible, as we showed for the $\{(3, 1), (9, 3)\}$ arrangement. However, other arrangements exhibit more than one possible intermediate configuration, as we discussed for the $\{(6, 2), (15, 5)\}$ arrangement. In all the cases, the number of islands in phase space is consistent with the Poincaré-Birkhoff fixed point theorem. However, the theorem does not determine the number of isochronous island chains, nor the transition between different configurations. For these features, we rely on the analysis of isochronous island chains in twist systems presented in Refs. \cite{Sousa_2013, sousa2014, sousa2015}.

In addition, for some limiter arrangements, we also identified shearless bifurcations inside isochronous islands. In this case, we calculated the local winding number profile with respect to the island center, and showed that it becomes nonmonotonic. Furthermore, a shearless invariant curve separates a pair of twin secondary island chains that appears inside the primary island. As the electric current $I_2$ increases, we also observed a separatrix reconnection process typical of nontwist systems occurring inside the islands of primary resonance. It is also important to point out that the local winding number changes from monotonic to nonmonotonic immediately before the primary islands undergo an isochronous bifurcation with the consequent creation of a new isochronous island chain. 

We point out that the complexity of physical phenomena is partially captured by dynamical systems and even further illuminated by a reduction to discrete maps, which emulate properties of simple systems. There has never been a question as to whether these maps portray quantitatively the discrete evolution on Poincaré sections. Instead, it is the qualitative features and their structural stability that matter. For instance, an anomalous bifurcation first detected in a map has a good chance to be present in a more realistic model for the system that motivated it, along with its unfolding with changing parameters. The point here is that, notwithstanding the auspicious eventuality that the new phenomena presented in this paper are detectable in the more involved models, it is quite possible that they would be missed without the foreknowledge of what and where to look for.

In this sense, our mapping results for a perturbed large aspect ratio tokamak indicate that similar analyses could be performed to interpret the heteroclinic bifurcations previously described for isochronous perturbations in the DIII-D and NSTX-U tokamaks.

\begin{acknowledgments}
We acknowledge financial support from CNEN (Comissão Nacional de Energia Nuclear), grant No 01341.001299/2021-54, the European Union’s Horizon 2020 research and innovation programme under the Marie Sk\l{}odowska-Curie grant agreement No 899987, São Paulo Research Foundation (FAPESP), Grant No 2018/03211–6, and Conselho Nacional de Desenvolvimento Científico e Tecnológico (CNPq), Grant No. 304616/2021-4.
\end{acknowledgments}

\appendix*
\renewcommand\thefigure{\thesection A\arabic{figure}}\setcounter{figure}{0}

\section{Arrangement $\mathbf{\{(3, 1), (9, 3)\}}$ - Separatrixes and bifurcations}

\begin{figure}[!tb]
    \centering
    \includegraphics[width=1.0\linewidth, height = 7.5cm]{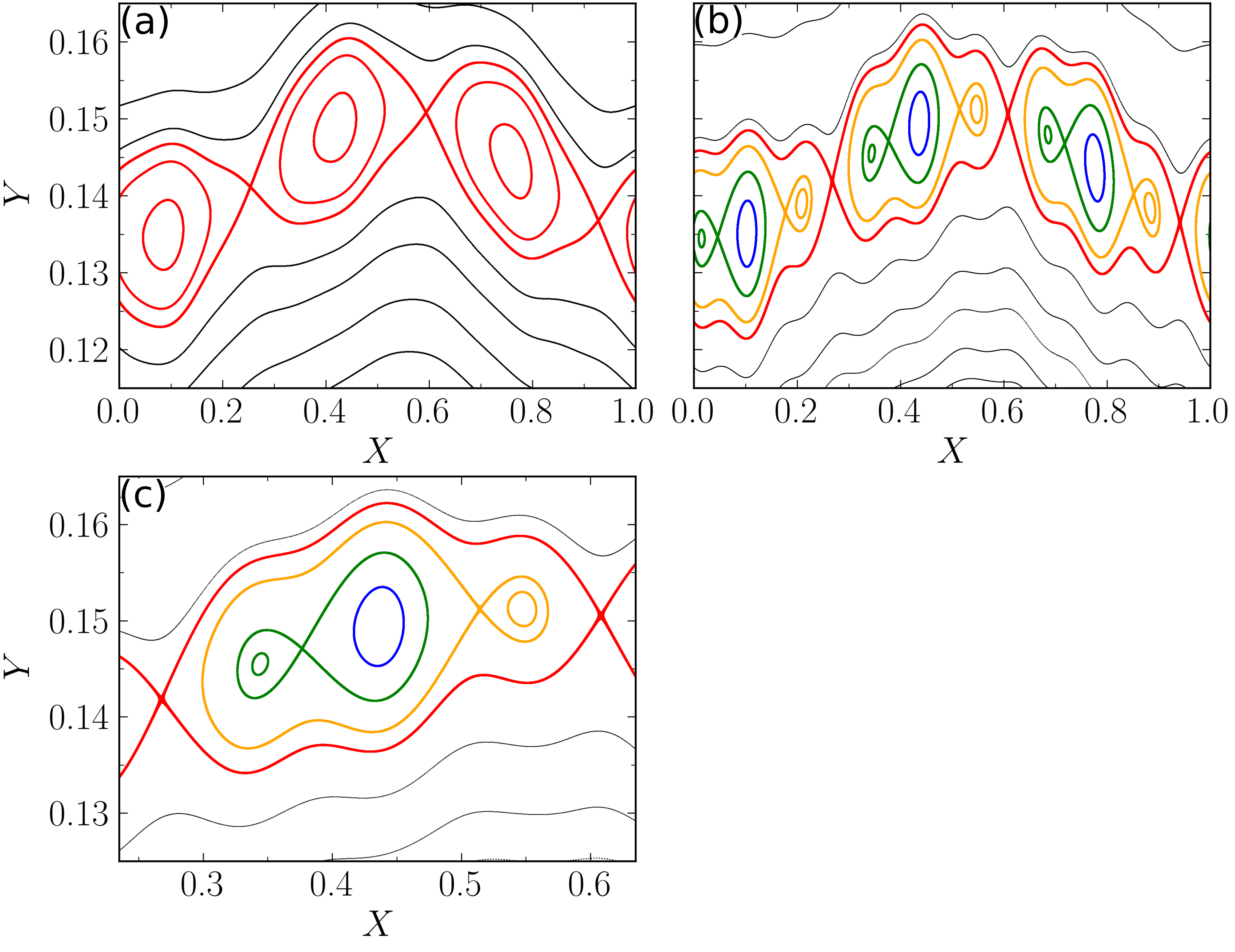}
    \caption{Poincaré sections for the $\{(3, 1), (9, 3)\}$ arrangement  with $\epsilon_1 = 1.13 \times 10^{-3}$ and a) $\epsilon_2 = 2.26 \times 10^{-4}$, and b) $\epsilon_2 = 1.694 \times 10^{-3}$. The isochronous islands correspond to those in Fig.~(\ref{arranjo_3_1_9_3}). But for these lower values of $\epsilon_1$ and $\epsilon_2$, it is possible to identify the islands separatrixes. Panel (c) is a zoom-in of panel (b).}    
    \label{appendix_arrangement_3_1_9_3_separatrixes}
\end{figure}

In Section \ref{Heteroclinic_Bifurcations}, the amplitudes of the limiter currents are chosen according to expected experimental values. However, for such currents the Poincaré sections shown in Fig.~(\ref{arranjo_3_1_9_3}) contain a significant area with chaos. To support our conclusions stated in Section \ref{Arragement_(3,1)_(9,3)}, we present Fig.~(\ref{appendix_arrangement_3_1_9_3_separatrixes})  obtained for smaller limiter electric currents, for which chaos in the Poincaré sections is still restricted to tiny areas. Thus, in Fig.~(\ref{appendix_arrangement_3_1_9_3_separatrixes}) we can identify the separatrixes of the isochronous islands corresponding to those displayed in Fig.~(\ref{arranjo_3_1_9_3}). We also notice that all the separatrixes surround more than one elliptic periodic point in Fig.~(\ref{appendix_arrangement_3_1_9_3_separatrixes}c).

\begin{figure}[!tb]
    \centering
    \includegraphics[width=1.0\linewidth,  height = 8cm]{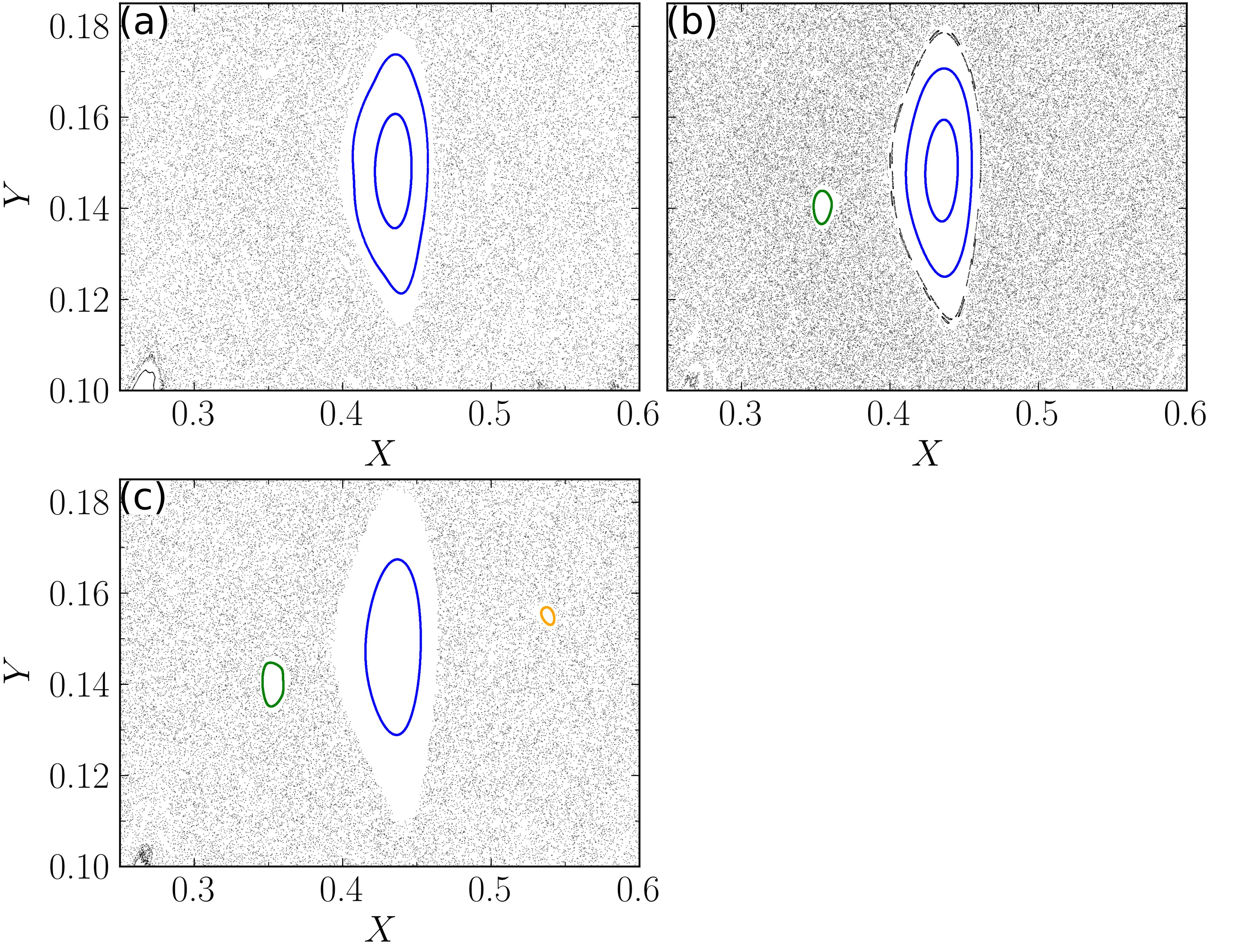}
    \caption{Poincaré sections amplification showing the onset of the small islands of Fig.~(\ref{ampliacao_3_1_9_3}) for the $\{(3, 1), (9, 3)\}$ arrangement with $\epsilon_1 = 4.4959 \times 10^{-2}$   and a) $\epsilon_2 = 4.1401 \times 10^{-2}$, b) $\epsilon_2 = 4.4620 \times 10^{-2}$, and c) $\epsilon_2 = 4.6315 \times 10^{-2}$.}    \label{appendix_arrangement_3_1_9_3_intermediate_mode_6_2}
\end{figure}

Still in Section \ref{Arragement_(3,1)_(9,3)}, we present Fig.~(\ref{ampliacao_3_1_9_3}) to discuss in more details the bifurcations for the $\{(3, 1), (9, 3)\}$ arrangement. Fig.~(\ref{appendix_arrangement_3_1_9_3_intermediate_mode_6_2}) shows that these saddle-node bifurcations have different thresholds for the onset of the two small island chains in green and yellow, as referred to in the analysis of Fig.~(\ref{ampliacao_3_1_9_3}). In other words, the green island to the left appears in the Poincaré section for lower values of $\epsilon_2$ than the yellow island to the right.

\bibliography{test}
\end{document}